\RequirePackage{rotating}
\documentclass[useAMS,usenatbib]{mn2e}
\usepackage{epsfig}
\usepackage{amssymb}
\usepackage{upgreek}
\usepackage{rotating}
\usepackage[utf8]{inputenc}
\usepackage{xcolor}
\usepackage{amsmath}

\usepackage{tabularx}

\newcommand\ion[2]{#1\,{\scshape{#2}}}

\title[Variable component of AGN spectra] {Reddening and the shape of the variable component of the continua of active galactic nuclei from the optical to the far ultraviolet. I.}

\author[C. Z. P. Heard and C. M. Gaskell]{Clio Z. P. Heard\thanks{E-mail:
czheard@ucsc.edu} and C. Martin Gaskell\thanks{E-mail:
mgaskell@ucsc.edu}\\
\\Department of Astronomy and Astrophysics, University of California, Santa Cruz, CA 95064
}

\begin{document}

\date{Received 2021 August 22; revised 2022 June 24}

\pagerange{\pageref{firstpage}--\pageref{lastpage}} \pubyear{2021}

\maketitle

\label{firstpage}

\begin{abstract}
We analyze the photometric variability of { 4,611} active galactic nuclei (AGNs) from the Sloan Digital Sky Survey Stripe 82. We recover the spectral energy distribution (SED) of the variable flux as a function of wavelength. For rest wavelengths longer than $\sim 2200$\AA\ we find that the SED  of the variable component of the bluest AGNs is {consistent with} the $F_{\nu} \propto \nu^{+1/3}$ spectrum predicted for an externally-illuminated accretion disc.  We confirm there is some residual variable emission corresponding to the ``small blue bump" and other broad-line region variability. We interpret steeper optical spectra of the variable component as being due to intrinsic reddening. This is supported by the correlation of the Balmer decrement with the colour excess of the variable component. We find the median {internal} reddening of SDSS AGNs in Stripe 82 with $z < 0.4$ to be $E(B-V) \thickapprox 0.10$ in agreement with the reddening derived from the Balmer decrement. {Individual AGNs in the sample can have $E(B-V) > 0.4$.}
\end{abstract}


\section{Introduction}

When analyzing the spectra of active galactic nuclei (AGNs), the attenuation by dust needs to be taken into account. Determination of the correct luminosity, spectral energy distribution, emission-line ratios of AGNs, and hence the physical conditions close to the central black hole, is not possible without allowance for attenuation from dust in the line of sight. Even though dust is an essential part of our standard model of AGNs (see \citealt{Antonucci93}), the quantity and location of obscuring dust in AGNs has long been controversial (see \citealt{Gaskell17} for a review). Our previous work \citep{Heard+Gaskell16} confirms that there is a substantial attenuation and implies that the dust causing the heaviest attenuation is located between the NLR and BLR.

Reddening can be determined either from known line ratios (e.g., hydrogen-line ratios -- see \citealt{Dong+08} and \citealt{Gaskell17}) or, if the true continuum shape is known, from broad-band colours . For AGNs, unfortunately, we do not know {\it a priori} what the intrinsic continuum shape is. However, \citet{Choloniewski81} made the important discovery that when AGNs vary, the colour of the variable component in the optical maintains the same shape.  Higher-quality observations have shown that, for a given object, the colour of the variable component is remarkably constant and that in the optical the constant component is consistent with being starlight from the host galaxy (see, for example, Figure 20 of \citealt{Sakata+10} or Figure 3 of \citealt{Ramolla+14}).  Cho{\l}oniewski proposed that when the colour of the variable component is redder, this is because of internal reddening in the AGN.  The Cho{\l}oniewski method has subsequently been used by \citet{Winkler+92} and \citet{Winkler97} to obtain reddening estimates for many AGNs.

In this paper we use the Cho{\l}oniewski method for a large-scale study of the continuum colours of the variable components of 4611 AGNs in order to investigate the reddening and continuum shape. This study  has two orders of magnitude more AGNs than previous studies using the Cho{\l}oniewski method to determine reddening, and, for the first time, includes the ultraviolet region of the spectrum.  In Paper II  (Heard \& Gaskell in preparation) we investigate the shape of the variable continuum in the UV as a function of luminosity, mass, and Eddington ratio.


\section{The Stripe 82 Sample and Method of Analysis}

A 290-square-degree equatorial region, known as ``Stripe 82", in the southern Galactic cap was repeatedly imaged by the Sloan Digital Sky Survey (SDSS) over a decade (see \citealt{Sesar+07} for details and \citealt{Ivezic+07}). {SDSS quasar candidates are selected by an automated algorithm via their nonstellar colors in $ugriz$ broadband photometry and also, for a smaller number of AGNs, by matching unresolved sources to the FIRST radio catalogs (see \citealt{Richards+02}).  The redshift distribution of the quasars in Stripe 82 is shown by \citet{Peth+11}.  } 

The variability of the thousands of AGNs in Stripe 82 has been the subject of several studies. \citet{MacLeod+10} modeled the time variability of quasars as a dampened random walk (as proposed by \citealt{Gaskell+Peterson87}). \citet{Palanque-Delabrouille+11} used variability in the $ugriz$ optical bands to identify quasars.  \citet{Meusinger+11}, \citet{Schmidt+12} and \citet{Kokubo+14} have studied the wavelength dependence of variability and its correlation with other properties.  \citet{Zuo+12} studied the correlations between optical variability and the physical parameters of quasars, looking at redshift, rest-frame wavelength, black hole mass, Eddington ratio, and bolometric luminosity. \citet{Andrae+13} measured the type-1 AGN luminosity function at $z=5$. \citet{Falomo+14} studied the host galaxies of low-redshift quasars. \citet{Hernitschek+15} estimated black hole masses of the AGNs.  \citet{LaMassa+16} used multi-wavelength (X-ray, optical and IR) to explore the efficiency of optical-IR diagnostics for finding hidden AGNs in X-ray surveys.

{While Stripe 82 has a width of only $\pm 1.26^{\mathrm o}$ in declination, it extends over $100^{\mathrm o}$ in RA.  Inspection of the Galactic reddening estimates, $E(B-V)_{\mathrm{Gal}}$, of \citet{Schlafly+Finkbeiner11} shows that while the Galactic reddening increases for the low Galactic latitudes at the ends of Stripe 82, the mean $E(B-V)_{\mathrm{Gal}}$ for the centre of the strip in the range $21^{\mathrm h}$ $20^{\mathrm m} <$ RA $< 02^{\mathrm h} 40^{\mathrm m}$ is $\sim 0.03$,  The scatter is only $\pm 0.01$ magnitudes, which is comparable to the uncertainties in the \citet{Schlafly+Finkbeiner11} reddening estimates.  We therefore excluded the ends of Stripe 82 at lower Galactic latitude and restrict our study to AGNS in the range $21^{\mathrm h}$ $20^{\mathrm m} <$ 
RA $< 02^{\mathrm h} 40^{\mathrm m}$. In this region the \citet{Schlafly+Finkbeiner11} reddenings are well approximated by a simple 5th order polynomial in RA.  We show our adopted Galactic reddenings in Table 1.}

We converted SDSS $ugriz$ magnitudes to fluxes $F_u$, $F_g$, $F_r$, $F_i$, and $F_z$ { in mJy (see \citealt{Stoughton+02})}. We made flux variability plots: $F_g$ vs $F_u$, $F_r$ vs $F_g$, $F_i$ vs $F_r$, and $F_z$ vs $F_i$. The slopes of the relationships between the fluxes for the different passbands were calculated.  We will refer to these as ``Flux variability gradients" (FVGs) {defined as the longer-wavelength change divided by the shorter-wavelength change. Thus a higher FVG corresponds to a redder variable continuum.}  These are shown in Figure 1. Because there are observational errors in both axes, we determined the gradients using the ordinary-least-square-bisector (``OLS-bisector") method of \citet{Isobe+90}. A small number of outlying points were identified by the Chauvenet criterion { (see \citealt{Taylor97})} and excluded from the analysis. We show sample plots in Figure 1.   Finally, {after correcting for Galactic extinction,} the FVGs were converted into local spectral indices, $\alpha$, {defined by $F_ {\nu} \propto \nu^{+\alpha}$} between the passbands. {The wavelengths of the $ugriz$ filters were taken to be $\uplambda$3543, $\uplambda$4770, $\uplambda$6231, $\uplambda$7625, and $\uplambda$9134 respectively.  Galactic extinction corrections were made using $E(u-g) = 1.06~E(B-V)$, $E(g-r) = 1.15~E(B-V)$, $E(r-i) = 0.66~ E(B-V)$, and $E(i-z) = 0.49~E(B-V)$}. The errors in the FVGs, and hence in $\alpha$, were calculated from the correlation coefficients. 

\begin{figure*}
  \includegraphics[width=0.8\textwidth]{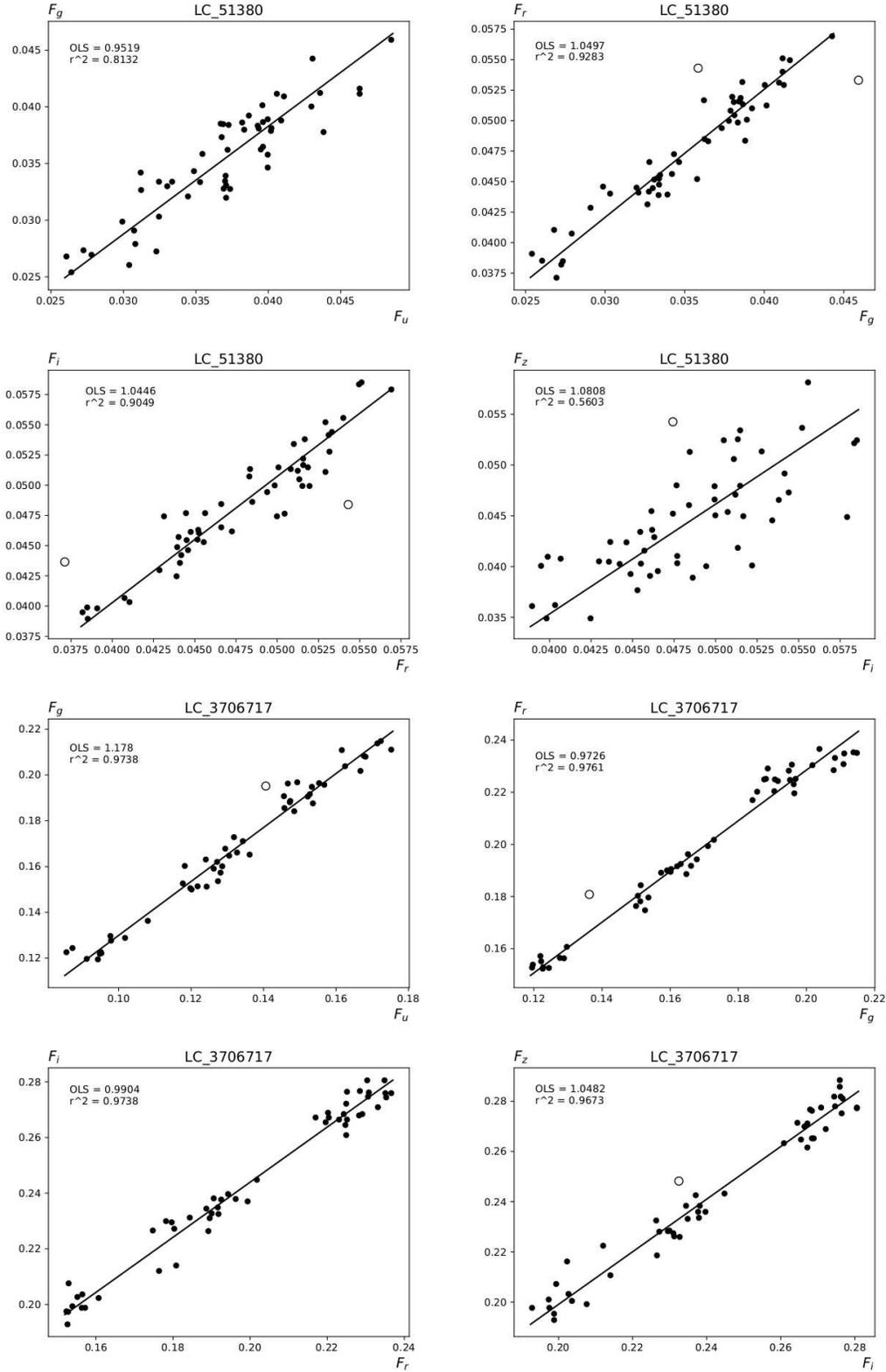}
  \caption{Sample flux-flux plots for two AGNs. The LC number refers to the light curve number in Stripe 82 (see \citealt{Ivezic+07}). { Fluxes are in mJy.} The open circle indicate outliers excluded in the analysis. The straight lines are the OLS-bisector fits. The flux-flux plots for the entire sample of 6306 AGNs are available online.}
\end{figure*}

\section{Results and Analysis}

\subsection{Flux-variability gradients and spectral indices}

Flux-variability plots were made for the fluxes of filters of adjacent wavelengths for all { 4611 AGNs}. Sample plots are shown in Figure 1 and the complete set is available on-line. The OLS-bisector lines are shown in Figure 1. The points excluded from the analysis are indicated in the plots but are not used in determining the FVGs. Inclusion or exclusion of these points has a negligible effect on the analysis. Table 1 gives the FVGs, { the adopted Galactic reddening,} the derived spectral indices, $\alpha$ (defined as
 $F_{\nu} \propto \nu^{+\alpha}$), and associated errors for all quantities. The table in its entirety is available online. The start of the table is given here to assist the reader.


\begin{sidewaystable*}


\scriptsize
\begin{tabularx}{1.0\textwidth} { 
  | >{\raggedleft}l
  | >{\centering}c
  | >{\raggedleft}l
  | >{\raggedleft}l
  | >{\raggedleft}l
  | >{\raggedleft}l
  | >{\raggedleft}l
  | >{\raggedleft}l 
  | >{\raggedleft}l
  | >{\raggedleft}l
  | >{\raggedleft}l
  | >{\raggedleft}l
  | >{\raggedright}r
  | >{\raggedleft}l
  | >{\raggedright}r
  | >{\raggedleft}l
  | >{\raggedright}r 
  | >{\raggedleft}l
  | >{\raggedright}r
  | >{\raggedleft\arraybackslash}l| }
  
  \hline

    &  &  &  & FVG & FVG & FVG & FVG & FVG & FVG & FVG & FVG & $\alpha$ \textit{(u-g)} & $\alpha$\textit{(u-g)} & $\alpha$\textit{(g-r)} & $\alpha$\textit{(g-r)} & $\alpha$\textit{(r-i)} & $\alpha$\textit{(r-i)} & $\alpha$\textit{(i-z)} & $\alpha$\textit{(i-z)}\\
  
   SDSS Name &  LC  & \textit{E(B-V)} & \textit{~~~z} & \textit{(u-g)} & \textit{(u-g)} & \textit{(g-r)} & \textit{(g-r)} & \textit{(r-i)} & \textit{(r-i)} & \textit{(i-z)} & \textit{(i-z)} &  & Err &  & Err &  & Err &  &  Err\\
     
   & &  & &  & Err &  & Err & & Err & & Err &  &  &  &  &  &  &  &\\
    \hline
   
J000006.53+003055.2 & 49159 & 0.0320 & 1.8177 & 0.993 & 0.070 & 0.817 & 0.039 & 0.973 & 0.060 & 1.535 & 0.164 & 0.128 & 0.255 & 0.882 & 0.174 & 0.232 & 0.319 & $-2.302$ & 0.606 \\			

J000008.13+001634.7 & 125164 & 0.0320 & 1.8359 & 1.115 & 0.109 & 1.075 & 0.062 & 1.366 & 0.093 & 1.391 & 0.164 & $-0.260$ & 0.354 & $-0.144$ & 0.210 & $-1.448$ & 0.352 & $-1.754$ & 0.669 \\

J000011.95+000225.3 & 24646 & 0.0320 & 0.4750 & 0.765 & 0.057 & 0.855 & 0.048 & 0.888 & 0.049 & 1.248 & 0.089 & 1.000 & 0.269 & 0.712 & 0.205 & 0.684 & 0.285 & $-1.151$ & 0.403 \\

J000012.25-003220.5 & 113067 & 0.0320 & 1.4394 & 0.979 & 0.057 & 0.949 & 0.031 & 0.988 & 0.048 & 1.316 & 0.139 & 0.175 & 0.210 & 0.322 & 0.119 & 0.156 & 0.251 & $-1.446$ & 0.599 \\

J000013.80-005446.7 & 35034 & 0.0320 & 1.8378 & 1.032 & 0.086 & 1.025 & 0.055 & 1.299 & 0.089 & 0.979 & 0.114 & $-0.001$ & 0.301 & 0.034 & 0.196 & $-1.199$ & 0.355 & 0.1980 & 0.660 \\

J000014.82-011030.7 & 18787 & 0.0320 & 1.8890 & 1.101 & 0.061 & 0.853 & 0.036 & 1.005 & 0.035 & 1.244 & 0.089 & $-0.217$ & 0.200 & 0.720 & 0.154 & 0.072 & 0.180 & $-1.133$ & 0.405 \\

J000015.47+005246.8 & 50026 & 0.0320 & 1.8461 & 1.184 & 0.039 & 0.835 & 0.017 & 0.905 & 0.030 & 1.049 & 0.071 & $-0.460$ & 0.119 & 0.800 & 0.074 & 0.590 & 0.171 & $-0.186$ & 0.383 \\

J000016.43-001833.3 & 2670 & 0.0320 & 0.7022 & 1.144 & 0.058 & 0.937 & 0.050 & 0.885 & 0.040 & 0.858 & 0.080 & $-0.345$ & 0.183 & 0.370 & 0.195 & 0.701 & 0.234 & 0.9320 & 0.528 \\

J000017.88+002612.6 & 36251 & 0.0320 & 0.5521 & 1.149 & 0.087 & 0.870 & 0.063 & 1.090 & 0.085 & 1.639 & 0.161 & $-0.360$ & 0.273 & 0.647 & 0.264 & $-0.330$ & 0.404 & $-2.666$ & 0.556 \\

J000030.37-002732.3 & 86582 & 0.0319 & 1.8043 & 1.245 & 0.075 & 0.979 & 0.045 & 0.969 & 0.049 & 1.199 & 0.114 & $-0.629$ & 0.217 & 0.205 & 0.168 & 0.252 & 0.262 & $-0.929$ & 0.538 \\

J000031.86+010305.2 & 68390 & 0.0319 & 1.0857 & 0.999 & 0.104 & 1.290 & 0.085 & 1.024 & 0.064 & 1.315 & 0.128 & 0.107 & 0.377 & $-0.825$ & 0.240 & $-0.021$ & 0.323 & $-1.442$ & 0.551 \\

J000032.70-005512.9 & 95334 & 0.0319 & 1.5183 & 0.960 & 0.063 & 0.967 & 0.044 & 1.021 & 0.066 & 1.044 & 0.111 & 0.241 & 0.237 & 0.252 & 0.166 & $-0.007$ & 0.335 & $-0.159$ & 0.603 \\

J000039.00-001803.9 & 62907 & 0.0319 & 2.1210 & 1.041 & 0.065 & 1.058 & 0.041 & 0.862 & 0.043 & 1.096 & 0.081 & $-0.030$ & 0.225 & $-0.085$ & 0.141 & 0.831 & 0.258 & $-0.430$ & 0.418 \\

J000042.02-004501.4 & 109616 & 0.0319 & 1.3118 & 0.800 & 0.053 & 0.963 & 0.045 & 0.969 & 0.046 & 1.203 & 0.109 & 0.850 & 0.239 & 0.267 & 0.170 & 0.252 & 0.246 & $-0.947$ & 0.513 \\			
															
J000042.90+005539.5 & 69839 & 0.0318 & 0.9450 & 1.037 & 0.034 & 1.090 & 0.028 & 0.989 & 0.023 & 1.101 & 0.041 & $-0.018$ & 0.118 & $-0.196$ & 0.094 & 0.151 & 0.120 & $-0.455$ & 0.210 \\			
J000046.15-003007.1 & 82001 & 0.0318 & 1.4414 & 1.158 & 0.041 & 0.888 & 0.017 & 1.007 & 0.024 & 1.001 & 0.064 & $-0.387$ & 0.128 & 0.570 & 0.070 & 0.061 & 0.123 & 0.0740 & 0.361 \\		
J000053.09-003712.7 & 42942 & 0.0318 & 1.3217 & 0.963 & 0.053 & 0.827 & 0.039 & 1.010 & 0.037 & 1.073 & 0.084 & 0.230 & 0.199 & 0.835 & 0.172 & 0.046 & 0.189 & $-0.312$ & 0.443 \\

  \hline
  
\end{tabularx}

~~\\

{Table 1.} Compilation of SDSS names, Light Curve numbers (LC) from \citet{Ivezic+07}, {Galactic reddenings}, redshift (\textit{z}), flux variability gradients (FVG), error of flux variability gradients, spectral indices ($\alpha$), and errors of spectral indices.
The complete table for all AGNs is available online. A portion of the table is shown here to assist in reading the on-line table.

\end{sidewaystable*}

\subsection{The spectral energy distribution of the variable component}

Figure 2 shows the distribution of local spectral indices against rest wavelength for the entire sample. Since there was photometry in five filters, there are usually four spectral indices per AGN.
\begin{figure*}
  \includegraphics[width=0.9\textwidth]{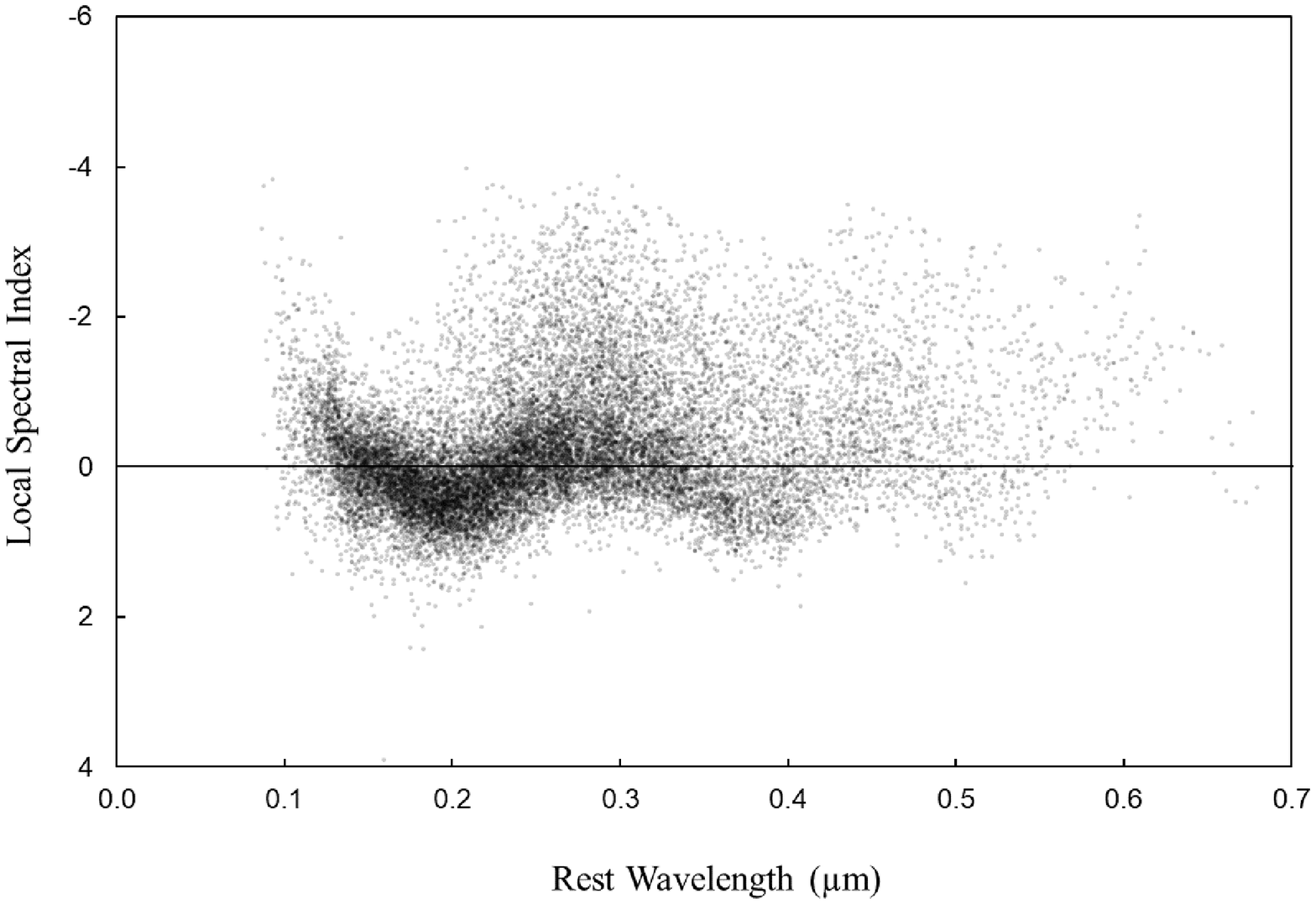}
  \caption{The local spectral indices, $\alpha$, as a function of rest-frame effective wavelength is the rest frame in microns for all four filter combinations for all 6306 AGNs. {The rest-frame effective wavelength is the average of the two filter wavelengths divided by $(1+z)$.} The position of a point above the lower envelope is interpreted as being due to reddening. Points have been made semi-transparent to make the image clearer.}
\end{figure*}
From Figure 2 it can be seen that (a) there is a fairly well defined minimum $\alpha$ at each rest wavelength. (i.e., the bluest AGNs at each rest wavelength), and (b) that this minimum $\alpha$ is not constant, but varies with wavelength. Following Cho{\l}oniewski, we interpret the lower envelope in Figure 2 as corresponding to the least reddened AGNs, and the position of a point above the lower envelope to be a function of reddening of that AGN.
{\citet{Mao+Zhang16} studied blazars in Stripe 82. Our sample includes none of their BL Lac objects, just one blazar, and two AGNs they classify as flat-spectrum radio quasars (FSRQs). Relative to other AGNs of the same wavelength, the blazar (J030458.97+000235.7) and one of the FRSQs (J032759.21+004422.7) are close to the average $\alpha$, while the second FSRQ (J211817.39+001316.8) is slightly redder than average.}

\subsection{Errors in the local spectral indices}

In determining the SED of the AGNs with the lowest reddening (i.e., the lower envelope in Figure 2) it is important to know the errors in our estimates of the spectral indices since the lower envelope is blurred by observational errors.  The errors in the FVGs and the corresponding errors in $\alpha$ are given by the OLS bisector fits and are given in Table 1.  We can get another indication of the errors in $\alpha$ by comparing the local spectral slopes deduced from different filter pairs.

The key assumption of the Cho{\l}oniewski method -- an assumption strongly supported by the linear flux-flux correlations -- is that the spectral shape of the {\em variable} component is constant {on the timescale of the variability}.  This shape can be well approximated over a wide spectral range by a power law. We therefore expect the local spectral indices at different wavelengths to be quite similar. In Figure 3, we show for AGNs with $0.35 < z < 0.4$, the relationship between the spectral indices $\alpha_{gr}$ determined from the $g$ and $r$ bands to $\alpha_{ri}$, the indices determined between the $r$ and $i$ bands. {We restrict the analysis in Figure 3 to $0.35 < z < 0.4$ to minimize the effects of changes in rest-frame wavelength.  This redshift range is also low enough to permit us to compare our reddening estimates from the continuum variability with estimates from the Balmer decrement (see below).} When comparing spectral indices derived from different filter pairs we expect a $45^{\circ}$ slope.  {The OLS-bisector fit in Figure 3 is in good agreement with this.  Reddening (see the red arrow in Figure 4) is expected to move points to the lower left almost parallel to this line. Scatter perpendicular to the line should only be due to observational errors while scatter along the line will be due to a combination of observational errors and intrinsic scatter in the observed shape of the spectrum. For a normal distribution of a variate one expects about 5\% of values to lie outside $\pm$ two standard deviations ($\pm 2\upsigma$).  In Figure 3 we show a circle with a radius of twice the average standard deviation. The number of points lying more than $2\upsigma$ away from the line in the perpendicular direction is consistent with the expected number. We next assumed that, in the absence of any observational errors, the distribution of reddened spectral indices would have a cutoff at the unreddened (bluest) value. The {\em observed} distribution would then be this distribution of reddened slopes convolved with the scatter due to measuring errors. To estimate the unreddened slope we adjusted the position of the centre of the circle so that the fraction of observed bluer points outside the circle (i.e., to the upper left in the diagram) was consistent with the number expected from a normal distribution.  This corresponds to the centre of the circle being at about the 25th percentile of bluest slopes.  Because of the small number of points, the uncertainty in the estimated unreddening spectral index in this redshift range is $\thickapprox \pm 0.15$}

\begin{figure}
 \centering \includegraphics[width = 9cm]{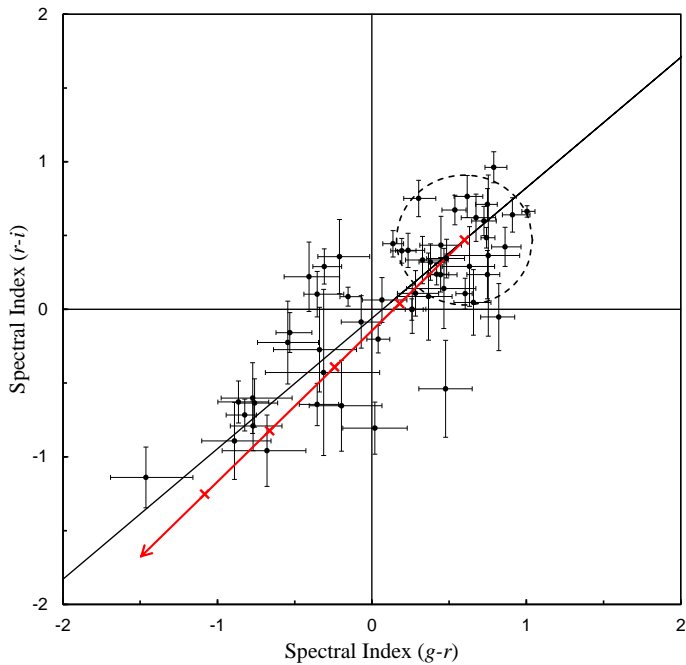}

 \caption{{Comparison of spectral indices $\alpha_{gr}$, from the $g$ and $r$ filters, with $\alpha_{ri}$, obtained from the $r$ and $i$ filters, for AGNs with redshift $0.35 < z < 0.4$.  The one standard deviation error bars are as given in Table 1 from the determination of the flux-variability gradients.  The diagonal line is an OLS-bisector fit.  The circle has a radius of twice the average error.  The centre of the circle is at the estimated unreddened spectral indices for the redshift interval (see text).  A reddening vector starting at the estimated unreddened spectral indices is shown by the red arrow.  The tick marks correspond to internal reddenings of $E(B-V)$ = 0.1, 0.2, 0.3, 0.4 and 0.5 magnitudes.  The reddenings are calculated from the mean AGN reddening curve of \citet{Gaskell+Benker07} for rest-frame wavelengths of the filters at $z = 0.375$.}}
\end{figure}

\subsection{Estimating the unreddened spectral energy distribution}
As noted, the lower envelope in Figure 2 is blurred by the observational errors in the spectral indices. {From the error analysis just discussed, we estimated that for our $0.35 < z < 0.4$ subsample, the unreddened slope is at about the 25th percentile of the distribution of slopes.  To estimate the lower envelope of the distribution of slopes at other redshifts we therefore} took the bluest {25\%} of slopes at each rest wavelength. {The percentile was calculated in bins of $\pm 100$ points and then resampled in wavelength intervals of 100 \AA.}

{The local spectral indices shown in Figure 2 show an obvious wave structure. This is readily explained by spectral features also varying with the continuum.  These produce broad bumps in the distribution of local spectral indices as a function of rest wavelength.  When there is extra emission, the effect is for $\upalpha$ to be greater (more positive) on the long wavelength side and less (more negative) on the short wavelength side.  If one averages over a wide wavelength interval, the average $\upalpha$ will be the $\upalpha$ of the underlying continuum.}  

{The continuum of an AGN rises to longer wavelengths because of emission from the hot dust in the infra-red (see, e.g., \citealt{Glass04}). This already affects the red end of the visible spectrum above 6000 \AA ~\citep{Winkler97} because dust near its sublimation temperature also emits in the optical \citep{Gaskell07}. The hot dust responds to continuum variability and so the effect of the dust will also be seen in the variability spectrum to the red.  Our study therefore focuses on wavelengths shorter than 6000 \AA ~in order to try to isolate the spectrum of the accretion disc.  In this paper we confine ourselves to wavelengths longer than $\sim \uplambda$2000 and defer discussion of what happens at shorter wavelengths to Paper II.}
We get a mean spectral index of {$+0.23$. As discussed, the statistical uncertainty of the method (see previous section) is $\thickapprox \pm 0.15$ in $\upalpha$. It should also be recognized that we are {\em assuming} that the bluest AGNs have zero reddening.  If it is not quite zero, the mean unreddened SED will be rising more steeply to shorter wavelengths.  The mean  spectral index of the bluest AGNs is thus consistent} with the prediction of $\upalpha = +0.33$ for an {\em externally-illuminated} accretion disc \citep{Friedjung85}\footnote{Note that the SED of an externally-illuminated accretion disc is the same as the well-known $F_{\nu} \propto \nu^{+1/3}$ spectrum predicted for the classic {\em internally}-heated accretion disc of \citet{Lynden-Bell69}.}. 

It is clear from Figure 2 that the lower envelope, contrary to what has been sometimes assumed, is \textit{not} a constant, but shows clear wavelength-dependent structure. {Integration of the 25th percentile bluest slopes gives the SED shown in Figure 4.  For reference we also show an $\alpha = +0.23$ power law.} The dominant deviation, as discussed by \citet{Kokubo+14}, is clearly the so-called ``small blue bump".  This bump is due primarily to a combination of blending of high-order Balmer lines into the Balmer continuum.  There is also a contribution from \ion{Fe}{ii} multiplets but \ion{Fe}{ii} emission is somewhat less variable than the high-order Balmer lines and continuum \citep{Gaskell+22}.  {The variability of the small blue bump contributes adds about 6\% to the variable SED above the rising power-law contribution.}

\begin{figure}
 \centering \includegraphics[width = 9
 cm]{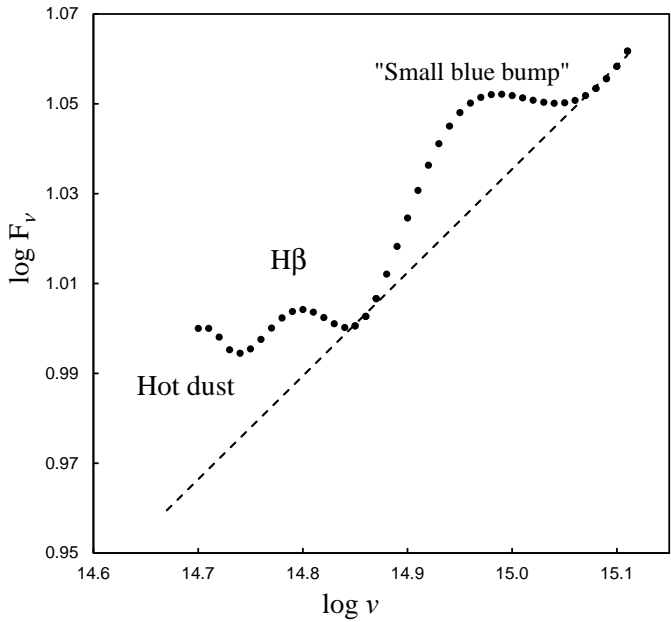}
 \caption{{The spectrum corresponding to the 25th-percentile bluest AGNs in Figure 2. The flux scale is arbitrary.  An $\alpha = +0.23$ power law is shown for reference. The locations of variable features not due to the accretion disc are indicated.}}
\end{figure}

The other obvious systematic deviation of the lower envelope in Figure 2 from a simple power law is $\lambda < 1700$\AA.  We defer discussion of this and the dependence of the SED on mass, luminosity, and black hole mass to Paper II.

\subsection{Reddening of Stripe 82 AGNs}


{We calculate $E(B-V)$ from $\Delta \alpha$, the difference between the local spectral index $\alpha$ and the 25th-percentile lower envelope of the distribution at the corresponding rest wavelength.} {$\Delta \alpha$ values for each filter pair $E(u-b), E(g-r)$ etc., were calculated using the equation:} 
\[
\Delta\upalpha = 0.4 E(\uplambda_{1},\uplambda_{2})/\log(\lambda_{2}/\lambda_{1})
\]

\noindent{Where $\uplambda_{1}$ and $\uplambda_{2}$ are the wavelengths of filters.}

 {The colour excesses in the Sloan filters were then converted into equivalent $E(B-V)$ values using the mean AGN reddening curve of \citet{Gaskell+Benker07} as follows.}

\begin{align*}
E(B-V) = E(u-g)[-0.2047 z^{3} + 0.283 z^{2} - 0.168 z + 1.2723]
\end{align*}

\begin{align*}
E(B-V) &= E(g-r)[0.0339 z^{4} -0.0277 z^{3} - 0.0592 z^{2} \\ &+ 0.0983 z + 1.1615]
\end{align*}

\begin{align*}E(B-V) &= E(r-i)[0.1107 z^{4} -0.2803 z^{3} + 0.1877 z^{2} \\ &+ 0.0833 z + 0.8738]
\end{align*}

\begin{align*}
E(B-V) &= E(i-z)[0.1678 z^{4} -0.5578 z^{3} + 0.6373 z^{2} \\& - 0.1615 z + 0.8201]
\end{align*}


{For each of the AGNs with a redshift $0.35 < z < 0.4$ we calculated the unweighted mean reddening of the three slopes from the four shortest wavelength filters. The slopes between the $i$ and $z$ filters were omitted because their errors were larger.  If the unreddened SED is taken to be the bluest 25\% (see above), the median intrinsic $E(B-V)$ (i.e., not including Galactic reddening) for the $0.35 < z < 0.4$ SDSS Stripe 82 AGNs is 0.10. If the unreddened SED is instead given by the bluest $10 \%$, then the median $E(B-V)$ is 0.07 magnitudes greater (i.e.,  $E(B-V) \thickapprox 0.17$.  It is obvious from Figures 2 and 4 that there is a long ``tail'' to much higher reddenings ($E(B-v) > 0.4$).  This is despite the bias of the colour selection of the SDSS to blue objects.  In Paper II we will discuss the question of luminosity dependence of reddening}

\section{Discussion}


If there is intrinsic reddening, we should see a correlation between the observed BLR Balmer decrement with the colour excess of the variable component, since the BLR and continuum originate within light-days/weeks of each other.  Such a correlation has already been found for 14 well-studied, nearby Seyfert galaxies (\citealt{Cackett+07} -- see their Figure 4).

\begin{figure}
 \centering \includegraphics[width = 8.3cm]{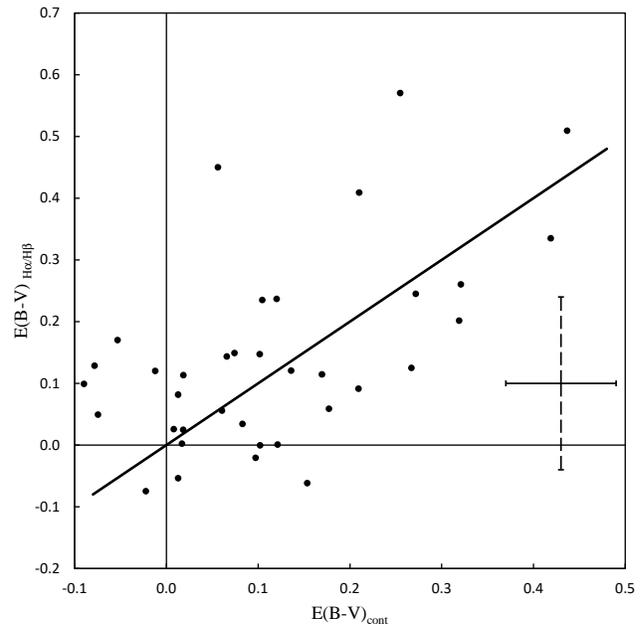}
 \caption{{Intrinsic} reddenings inferred from the broad-line Balmer decrements given in SDSS DR7 (see text for details) versus the average reddenings from the $(u-g), (g-r)$ and $(r-i)$ colours of the variable AGN continuum. {Both axes are corrected for Galactic reddening.} A typical error bar (see text) is shown in the lower right.  { The vertical error bar is shown with a dashed line to indicate that it is an average estimated minimum error. The diagonal corresponds to the reddening estimates being the same.}}
\end{figure}


In Figure 5 we plot the {intrinsic} reddenings inferred from the BLR Balmer decrements { for the low-redshift AGNs ($z < 0.4$)} against the {intrinsic} reddenings we deduce from the colour variability.  For reddenings from the flux variability we took the averages of the reddenings from the $(u-g), (g-r)$ and $(r-i)$ colours. The BLR Balmer decrements are based on SDSS DR7. \citet{Gaskell17} inferred the typical unreddened Balmer decrement for an integrated line profile using the line fluxes of \citet{Dong+08}. Because there are systematic differences in estimates of line intensities by different groups, we have scaled the SDSS DR7 Balmer decrements to the \citet{Dong+08} Balmer decrements by multiplying the former by 0.72 {because the \citet{Gaskell17} estimate of the unreddened Balmer decrement used the \citet{Dong+08} measurements}.  We then calculated the BLR reddenings shown on the vertical axis in Figure 5 using the unreddened Balmer decrement of H$\upalpha$/H$\upbeta = 2.7$ derived by \citet{Gaskell17} from the \citet{Dong+08} line fluxes measurements.   Different choices of scaling and unreddened H$\upalpha$/H$\upbeta$ simply move all points up and down in Figure 5.

As can be seen, Figure 5 shows that there is indeed a significant correlation ($p = 0.0004$) between the two reddening estimates, although with a large scatter.  We show a typical error bar at the bottom right of the figure. The typical error bar in $E(B-V)_{\mathrm{cont}}$ of $\pm 0.06$ was calculated from the dispersion in the reddenings from the three filter combinations.  We estimated a {\em minimum} typical error bar in $E(B-V)_{\mathrm{H\upalpha/H\upbeta}}$ by intercomparing the estimates from the  H$\upalpha$/H$\upbeta$ ratios in DR7 with those in \citet{Dong+08}.  This gives an RMS dispersion between the two measurements of $E(B-V)$ of $\pm 0.14$.  If we assume that the errors are in the two sources this gives a typical error of $\pm 0.10$.  However, the two sources are measuring {\em the same spectra} so this error estimate only gives the uncertainties due to the measuring techniques.  Hence we say that this is a minimum error. Since this is already almost twice the error in $E(B-V)_{\mathrm{cont}}$, this shows that much of the scatter in Figure 5 is primarily due to the difficulty of determining $E(B-V)$ from the Balmer decrement. In addition there can be intrinsic scatter. For any given AGN, the reddenings of the BLR and of the optical continuum do not have to be exactly the same since the dust and gas above the accretion disc and BLR can be patchy (see discussion in \citealt{Gaskell+Harrington18} and \citealt{Jaffarian+Gaskell20}).   



The mean slope of the unreddened variable component over a wide range of wavelength is consistent with the $\alpha = +0.33$ slope of an externally-illuminated accretion disc. The clear departures from this (see Figures 2 and 4) have natural explanations. As mentioned, the largest one, around $0.25 - 0.3$ microns, corresponds to the small blue bump. As is well-known, the BLR lags the continuum variability by light-days to light-months or longer. The flux-variability diagrams are clearly picking up not just the short-timescale continuum variability, but also the longer-timescale variability of the BLR lines, bound-free continua {and hot dust}. This needs to be allowed for when using the Cho{\l}oniewski method to determine reddenings.

\section{Conclusions}

From the study of the shape of the variable component of the continuum, we conclude that:

\begin{enumerate}
\item The unreddened spectral energy distribution of the variable component of the continuum of AGNs {rises to shorter wavelengths and is consistent with} the $F_{\nu} \propto \nu^{+1/3}$ spectrum expected from an externally-illuminated accretion disc.

\item The {optical to near-UV}variable component of the SED also includes a contribution at about the 6\% level from the BLR Balmer continuum and UV \ion{Fe}{ii} emission making up the small blue bump. This, and weaker contributions due to other BLR emissions such as H$\upalpha$, need to be allowed for in using the Cho{\l}oniewski method to determine reddenings for individual objects.

\item {The median internal reddening of SDSS AGNs is $E(B-V) \thickapprox 0.10$ but there is a long tail to the distribution towards higher reddenings.  For some of the AGNs $E(B-V) > 0.4.$}

\end{enumerate}

\section{Data Availability}

The data underlying this article are available in the online supplementary table.

\section*{Acknowledgments}

We are grateful to the referee, Harmut Winkler, for very helpful suggestions and careful reading of the paper. We also wish to thank Ski Antonucci for comments and discussion.


\end{document}